\documentclass[12pt,a4paper]{article}

\usepackage{amsmath,amsfonts,latexsym,amssymb}
\usepackage{verbatim,amsthm,curves,graphics}
\usepackage{mathrsfs}

\theoremstyle{definition}
\newtheorem{thm}{Theorem}[section]
\newtheorem{lemma}{Lemma}[section]

\newtheorem{defn}{Definition}[section]

\begin{document}

\newcommand{\mr}{\mathbb{R}} 
\newcommand{\mz}{\mathbb{Z}} 
\newcommand{\mc}{\mathbb{C}} 
\newcommand{\mh}{\mathbb{H}} 
\newcommand{\ms}{\mathbb{S}} 
\newcommand{\mn}{\mathbb{N}} 
\newcommand{\pgator}{/\!\!\! S}
\newcommand{\rP}{\operatorname{P}}
\newcommand{\g}{{\bf g}}
\newcommand{\car}{\operatorname{CAR}}
\newcommand{\rSpin}{\operatorname{Spin}}  
\newcommand{\rSO}{\operatorname{SO}}      
\newcommand{\rO}{\operatorname{O}}        
\newcommand{\rCliff}{\operatorname{Cliff}}
\newcommand{\WF}{\operatorname{WF}}       
\newcommand{\Pol}{{\rm WF}_{pol}}         
\newcommand{\cO}{\mathcal{O}}             
\newcommand{\cA}{\frak{A}}                
\newcommand{\fF}{\frak{A}}
\newcommand{\cV}{\mathscr{V}}             
\newcommand{\cR}{\mathscr{R}}             
\newcommand{\cC}{\mathscr{C}}             
\newcommand{\cF}{\mathscr{F}}             
\newcommand{\cK}{\mathscr{K}}             
\newcommand{\cH}{\mathscr{H}}
\newcommand{\cU}{\mathscr{U}}
\newcommand{\cB}{\mathscr{B}}
\newcommand{\cT}{\mathscr{T}}
\newcommand{\cN}{\mathscr{N}}
\newcommand{\cD}{\mathscr{D}}             
\newcommand{\cL}{\mathscr{L}}             
\newcommand{\cQ}{\mathscr{Q}}             
\newcommand{\cI}{\mathcal{I}}             
\newcommand{\bS}{\operatorname{S}}
\newcommand{\bU}{\mathcal{U}}
\newcommand{\bL}{\operatorname{OP}}
\newcommand{\glob}{{\textrm{\small global}}}
\newcommand{\loca}{{\textrm{\small local}}}
\newcommand{\singsupp}{\operatorname{singsupp}}
\newcommand{\dom}{\operatorname{dom}}
\newcommand{\clo}{\operatorname{clo}}
\newcommand{\supp}{\operatorname{supp}}
\newcommand{\rd}{{\rm d}}                 
\newcommand{\mslash}{/\!\!\!}             
\newcommand{\slom}{/\!\!\!\omega}         
\newcommand{\dirac}{/\!\!\!\nabla}        
\newcommand{\myid}{\leavevmode\hbox{\rm\small1\kern-3.8pt\normalsize1}}
\newcommand{\esssup}{\operatorname*{ess.sup}}
\newcommand{\ran}{\operatorname{ran}}
\newcommand{\sd}{{\rm sd}}
\newcommand{\reg}{\,{\rm l.n.o.}}
\newcommand{\wx}{{\bf x}}
\newcommand{\wk}{{\bf k}}
\newcommand{\ws}{{\bf s}}
\newcommand{\lno}{:\!}
\newcommand{\rno}{\!:}
\newcommand{\clim}{\operatorname*{coin.lim}}
\renewcommand{\min}{{\textrm{\small int}\,\mathscr{L}}}
\renewcommand{\max}{\operatorname{max}}
\renewcommand{\Im}{\operatorname{Im}}
\renewcommand{\Re}{\operatorname{Re}}
\newcommand{\Exp}{\operatorname{Exp}}
\newcommand{\Ad}{\operatorname{Ad}}
\newcommand{\sa}{\mathfrak{sa} } 
\newcommand{\so}{\mathfrak{so} } 
\renewcommand{\o}{\mathfrak{o} } 
\newcommand{\su}{\mathfrak{su} } 
\newcommand{\sq}{\mathfrak{sq} } 
\renewcommand{\sp}{\mathfrak{sp} } 
\renewcommand{\sl}{\mathfrak{sl}} 
\newcommand{\gl}{\mathfrak{gl}} 
\newcommand{\h}{\mathfrak{h} } 
\newcommand{\f}{\mathfrak{f} } 
\newcommand{\p}{\mathfrak{p} } 
\newcommand{\U}{\operatorname{U}} 
\newcommand{\Z}{\operatorname{Z}} 
\newcommand{\SO}{\operatorname{SO} } 
\newcommand{\SU}{\operatorname{SU} } 
\renewcommand{\O}{\operatorname{O} }
\newcommand{\SP}{\operatorname{Sp} } 
\newcommand{\SL}{\operatorname{SL}} 
\newcommand{\GL}{\operatorname{GL}} 
\newcommand{\Der}{\operatorname{Der}} 
\newcommand{\Str}{\operatorname{Str}} 
\newcommand{\End}{\operatorname{End}} 
\newcommand{\Cl}{\operatorname{Cl}} 
\newcommand{\ds}{\dot{+}}
\renewcommand{\aa}{\alpha}
\renewcommand{\gg}{\gamma}
\renewcommand{\d}{\delta }
\newcommand{\x}{\bar{x} } 
\newcommand{\y}{\bar{y} }
\newcommand{\s}{\bar{s} }
\newcommand{\diag}{\operatorname{diag}}
\newcommand{\ad}{\operatorname{ad}}
\newcommand{\sgn}{\operatorname{sgn}}
\newcommand{\tr}{\operatorname{tr}} 
\newcommand{\per}{\operatorname{per}} 
\newcommand{\Tri}{\operatorname{Tri}}

\title{
\rightline{\small PACS numbers: 11.10.Cd, 10.30.Ly and 11.10.Gh}
\vspace{1cm}
Noether Charges for Self-interacting
Quantum Field Theories in Curved Spacetimes with a Killing-vector}
\author{Stefan Hollands\footnote{Electronic mail:
        \tt stefan@bert.uchicago.edu}\hspace{0.5em}\\
       \it{Enrico Fermi Institute, Univ. of Chicago, 5640 Ellis Avenue,} \\ 
       \it{Chicago, IL 60637 - 1433, USA}}
\date{\today}

\maketitle

\begin{abstract}
We consider a self-interacting, perturbative 
Klein-Gordon quantum field in a curved spacetime 
admitting  a Killing vector field. We show that the action of 
this spacetime symmetry on interacting field operators can 
be implemented by a Noether charge which arises, in a certain 
sense, as a surface integral over the time-component of 
some interacting Noether current-density associated with the Killing field. 
The proof of this involves the demonstration of a corresponding 
set of Ward identities. Our work is based on the 
perturbative construction by Brunetti and Fredenhagen 
(Commun.Math.Phys. 208 (2000) 623-661)  
of self-interacting quantum field theories in general globally 
hyperbolic spacetimes. 
\end{abstract}

\vspace{1cm}
\rightline{\small KEYWORDS: Quantum field theory, perturbation theory, Noether charges, Killing vectors}

\pagebreak

\section{Intorduction}

It is widely believed that the theory of quantum fields in 
a classical curved spacetime background (for an introduction to this 
subject, see e.g.~\cite{wald,kaywald}) 
provides an excellent description 
of physical processes in which both gravitational effects
as well as the quantum nature of matter on the microscopic 
level are important, but in which quantum effects 
of the gravitational field and back-reaction are not 
important. Among others, it led to the spectacular prediction~\cite{hawking}, 
by Hawking in 1975, that black holes ``are not black'', but 
instead emit thermal radiation at temperature 
$T_H = \frac{\kappa}{2\pi}$, where $\kappa$ is the surface 
gravity of the horizon.
Moreover, the analysis of the 
conceptual basis of quantum field theory (QFT) in curved spacetime 
has also lead to a reassessment 
of the general structure of QFT, since
it forces one to emphasize the principle of locality, and 
abandon concepts such as preferred global vacuum states, the 
conventional notion of ``particle'', unitary time-evolution, etc.  

Most effects predicted by QFT in curved spacetime can already be seen in linear
(i.e., non self-interacting) models, and in fact the vast majority 
of the work done in the subject has been devoted to such models.
However, they are not realistic and one therefore also ought 
to consider self-interacting quantum fields on curved backgrounds
(of course one expects that the essential physical effects predicted
by linear models remain). Unfortunately, it does not seem to be 
clear at present---even in Minkowski space---just what 
a self-interacting quantum field really is, phrased in sensible 
mathematical terms, exept in some rather special 
models in low spacetime dimensions. On the other hand, it 
has long been known, at least in Minkowski space,  
how to construct self-interacting quantum field theory models on the level 
of perturbation theory. The challenge to construct perturbatively
also self-interacting quantum fields in curved spacetime has 
been taken up by several authors~\cite{bunch1,
bunch2,bf}, most recently\footnote{
We note that, while
\cite{bf} applies to any smooth, globally hyperbolic 
spacetime, this is not the case for the construction by \cite{bunch1,bunch2}, who
approach the problem from the point of view of Euclidean 
spaces rather than Lorentzian spacetimes. 
Now, if a Lorentzian spacetime can be viewed as a real
section of some complex space which also posses a real Euclidean
section, then the quantities of interest for the interacting 
theory in that Lorentzian spacetime 
can be obtained, via analytic continuation, from suitably defined 
corresponding quantities in the Euclidean section. 
However, apart from the static ones, there are essentially no curved 
spacetimes with this property and we do not want to restrict ourselves
to such a limited class of spacetimes here.}
by \cite{bf}, who have 
developed a variant of the causal approach to perturbation 
theory~\cite{eg,stuck,BS,S} for curved spacetimes, thereby arriving
at the same classification of theories into (perturbatively)
renormalizable and non-renormalizable ones as in Minkowski space.\footnote{
We must point out that the finite renormalization ambiguities found in 
\cite{bf}, even for renormalizable theories, are much greater than for the
corresponding theory in Minkowski space: Instead of free parameters 
they consist of free functions. This problem has been resolved in~\cite{howa}, but 
for the purposes of this article we can ignore this issue. 
}

One interesting result of \cite{bf} (which seems to be new even in 
flat spacetime) is that it is always possible construct algebras
$\cA_\cL(\cO)$ of interacting quantum fields 
(given by formal power series in the coupling strength) 
localized in some bounded 
region $\cO$ in spacetime, where the subscript 
$\cL$ indicates that the theory is defined by a local 
interaction Lagrangian $\cL$ (the construction of these 
algebras will be reviewed in some detail in Sec.~\ref{sec1}). 
These local algebras have the 
property that $\cA_\cL(\cO_1) \subset \cA_\cL(\cO_2)$ if
$\cO_1 \subset \cO_2$, and one can therefore define the algebra
of all observables of the theory as the inductive limit
$\cA_\cL = \cup_{\cO} \cA_\cL(\cO)$. 
This is quite remarkable, because it means that one can 
perturbatively construct the theory on the level of observables 
even in cases where an $S$-matrix does not properly exist 
because of incurable infrared divergences. 

In this work we want to study a self interacting Klein-Gordon
field on globally hyperbolic spacetimes $(M, \g_{ab})$ which
have an everywhere non-vanishing Killing vectorfield $\xi^a$
and for which there exists a $\xi^a$-invariant quasifree 
Hadamard state (an explanation of this term will be given below)
for the corresponding linear quantum field\footnote{
We remark that, since all our constructions are local, 
all the results in this paper still hold 
under the weaker assumption that there exists a 
$\xi^a$-invariant Hadamard state for every globally
hyperbolic set $\cO$ in $M$ with compact closure. Hence
the assumption that there is a {\it globally} defined
$\xi^a$-invariant quasifree Hadamard state (which might 
not exist, due to infrared problems) is not essential.}.
This class of spacetimes includes many examples of 
physical interest, for example the exterior of the 
Schwarzschild solution to Einstein's equation. 
Under these conditions, there 
exists an action of the corresponding 1-parameter symmetry group 
of $M$ by automorphsims $\alpha_{\cL, t}$ on the 
algebra of interacting fields $\cA_\cL$, such that 
$\alpha_{\cL, t}(\cA_\cL(\cO)) = \cA_\cL(\psi^t\cO)$, 
where $\psi^t$ is the flow of $\xi^a$. 

Now, it is a well-established feature of QFT in Minkowski spacetime
that the (infinitesimal) action of a 1-parameter family of 
(unbroken) symmetries of a theory can 
be implemented, at least locally, by a charge operator which arises
as a surface integral of a corresponding conserved Noether current-density. 
One expects that this should also hold in the present situation, 
where the 1-parameter group of symmetries in question now arises from  
the Killing vector field of the underlying curved spacetime. 
The aim of this paper is to show that 
this expectation is indeed correct. Namely, we show that for any 
bounded region $\cO$ and any globally hyperbolic region $\widehat \cO$
``strongly containing'' $\cO$, 
there is an operator $Q_\cL \in \cA_\cL(\widehat \cO)$
such that the infinitesimal action of $\alpha_{\cL, t}$ 
on any element $a \in \cA_\cL(\cO)$ is given by $[Q_\cL, a]$
(we mean the commutator). 
Moreover, $Q_{\cL}$ arises, in a certain sense, as a surface integral of 
the time-component of a covariantly 
conserved interacting Noether current-density 
$J^a_\cL$ over an arbitrarily chosen Cauchy surface $\widehat\Sigma$ of 
$\widehat \cO$. (We say that $\widehat \cO$ strongly contains $\cO$
if there is a Cauchy-surface $\widehat \Sigma$ of $\widehat
\cO$ and an open set $U \subset \widehat \Sigma$ such 
that the closure of $J(\cO) \cap \widehat \Sigma$ is 
contained in $U$.)

The main part of our proof of this result consists in showing that 
the perturbatively defined interacting current $J^a_\cL$ is 
covariantly conserved, which in turn is equivalent to the validity
of a corresponding set of Ward identities between certain time ordered products. 
A general framework to prove Ward identities within the causal approach was developed by 
D\"utsch and Fredenhagen~\cite{df} (they treated the case of 
the global $U(1)$-currents for QED in Minkowski spacetime), we here 
adopt their strategy. The crucial step in the proof is to remove 
the anomaly. In the case studied by~\cite{df} (and similarly 
\cite{pr,ho}, where the Ward identities for the interacting stress 
tensor in Minkowski space are treated), this is done by an 
argument based on momentum space techniques. 
These are not available in the case studied here, 
and one therefore has to proceed by an entirely different 
argument. 

We remark that, while we explitely deal only with a scalar self-interacting
field, a generalization of our results to other types of fields should be
straightforward.   

Our plan for this paper is as follows: In Sec.~\ref{sec1} we review 
the quantization of a linear Klein-Gordon field on a curved 
spacetime. After that, we recall the notion of Wick products
in curved space and then explain in some detail the 
construction of interacting fields, the algebras $\cA_\cL(\cO)$,
etc. in curved spacetime (here we follow~\cite{bf}). 
In Sec.~\ref{sec2}, we define the automorphic action 
$\alpha_{\cL,t}$ of the group of spacetime symmetries 
on the observables in the interacting theory. We then
show that this action is implemented by a local charge 
operator $Q_\cL$. The Ward identities for 
$J^a_\cL$ are proved in the Appendix.

\medskip

{\it Notations and conventions:} $(M, \g_{ab})$ denotes a 
globally hyperbolic, time-oriented  four dimensional curved spacetime of signature $+2$. 
$J^\pm(\cO)$ is the causal future 
respectively past of a subset $\cO \subset M$, and $J(\cO) = J^+(\cO) \cup J^-(\cO)$. 
By $D(\cO)$ we mean the domain of dependence of a region $\cO \subset M$, 
defined as the set of all points $x$ such that every future or every past
directed inextentible causal curve starting at $x$ intersects 
$\cO$. The metric volume element is denoted by $\mu_\g$ 
and the wave operator in curved spacetime by 
$\square_\g = \g^{ab}\nabla_a\nabla_b$. $\cD(M)$ denotes 
the space of compactly supported (complex-valued) testfunctions
on $M$ and $\cD'(M)$ is the corresponding dual space of 
distributions. We shall also find it useful 
to use a multi-index notation. If $\alpha = (\alpha_1, \dots, \alpha_n)$
is a multi-index, then we set $\alpha! = \prod_i \alpha_i!$.

\section{Review of QFT in CST}
\label{sec1}
\subsection{Quantization of linear scalar fields}

The theory of a free quantized Klein-Gordon field in curved spacetime
can be formulated in various ways. For our purposes, it is 
essential to formulate the theory within the 
so-called ``algebraic approach'' (see, for example \cite{wald,kaywald}). 
In this approach, one starts from an abstract *-algebra 
$\cA$ (with unit), which is generated by certain 
expressions in the smeared quantum field, $\Phi(f)$, where $f$ is
a test function. In \cite{wald,kaywald}, expressions of the form $e^{i\Phi(f)}$
were considered. The main advantage of working with such expressions is 
that the so-obtained algebra then has a norm (in technical terms, it is 
a $C^*$-algebra). Defining the algebra $\cA$ in that way would however be 
inconvenient for our purposes. 
Instead, we shall take $\cA$ to be the *-algebra generated by 
the identity and the smeared field operators 
$\Phi(f)$ themselves, subject to the following relations:

\medskip
\noindent
{\bf Linearity:} The map $f \mapsto \Phi(f)$ is complex linear.\\
{\bf Klein-Gordon Equation:} $\Phi((\square_\g - m^2)f) = 0$ for all 
$f \in \cD(M)$.\\ 
{\bf Hermiticity:} $\Phi(f)^* = \Phi(\bar f)$ for all $f \in \cD(M)$.\\  
{\bf Commutation relations:}
$[\Phi(f_1), \Phi(f_2)] = i\Delta(f_1, f_2)\myid$ for all 
$f_1, f_2 \in \cD(M)$, where $\Delta$ is the 
causal propagator (commutator function), defined as 
the difference $\Delta_A - \Delta_R$ between the uniquely determined 
advanced and retarded fundamental solutions of the Klein-Gordon
operator  $\square_\g - m^2$.  

\medskip
\noindent
The so-obtained algebra 
$\cA(M, \g)$ is now no 
longer a $C^*$-algebra, because of the unbounded 
nature of the smeared quantum fields $\varphi(f)$. This will however not 
be relevant in the following. We will often use the informal notation
\begin{equation*}
\Phi(f) = \int_M \Phi(x) f(x) \, {\bf \mu}_\g(x),
\end{equation*}
for the smeared quantum fields, and also for other distributions.
The local algebras $\cA(\cO)$ are by definition
the subalgebras of $\cA$ generated
by quantum fields smeared with testfunctions supported in $\cO \subset M$. 
Clearly these local algebras fulfill isotony, $\cA(\cO_1) \subset \cA(\cO_2)$ 
if $\cO_1 \subset \cO_2$. Since $\Delta(x_1, x_2) = 0$ for spacelike 
related points, the local algebras also satisfy spacelike 
commutativity, $[\cA(\cO_1), 
\cA(\cO_2)] = \{0\}$ if $\cO_1$ and $\cO_2$ are spacelike. 

A state in the algebraic framework is by definition a positive, 
normalized linear functional on $\omega: \cA \to \mc$, i.e., a linear functional
with the properties $\omega(a^*a) \ge 0$ for all $a \in \cA$ 
and $\omega(\myid) = 1$. The algebraic notion of state is related to 
the usual Hilbert-space notion of state via the so-called 
``GNS--theorem'', which says that for any algebraic
state $\omega$, there exists a representation $\pi_\omega$ of $\cA$ on 
a Hilbert space $\cH_\omega$, and a cyclic vector $|\Omega_\omega\rangle$ 
such that $\omega(a) = \langle \Omega_\omega| 
\pi_\omega(a) \Omega_\omega \rangle$ for all $a \in \cA$. 
If the state in question had been obtained from some set 
$u_\lambda$ of positive frequency solutions to the Klein-Gordon
equation, then the GNS--construction gives the usual representation 
of the field operators on Fock space (with $|\Omega_\omega\rangle$ the 
Fock vacuum), 
$\Phi(x) = 
\sum_\lambda a_\lambda^{\,} u_\lambda^{\,}(x) + a_\lambda^+ u_\lambda^+(x)$, 
where $a^+_\lambda$ are the creation operators corresponding to the 
modes $u_\lambda$.
The multilinear functionals on $\cD(M)$ defined by 
\begin{equation*}
\omega^{(n)}(f_1, \dots, f_n) = \omega(\Phi(f_1) \dots \Phi(f_n))
\end{equation*}
are called ``$n$-point functions'' (of the state $\omega$). 
A state is called ``quasifree''
if it has a vanishing one-point function and  
vanishing truncated $n$-point functions for $n > 2$. 
Prime examples for quasifree states are the states obtained from 
some set of positive frequency solutions $u_\lambda$. 
A state is called ``globally Hadamard'' if its two-point function
has no spacelike singularities and if its symmetrized two-point function
locally has the form of a Hadamard fundamental solution, $H$, 
given by 
\begin{eqnarray*}
H(x_1, x_2) = 
u(x_1, x_2) \, {\rm P}(\sigma^{-1}) + v(x_1, x_2) \ln |\sigma| + w(x_1, x_2).
\end{eqnarray*}
Here, $\sigma$ is the signed, squared geodesic distance between $x_1$ and
$x_2$, $u$ and $v$ are certain smooth, symmetric functions constructed 
from the metric, ``P'' means the principal value, 
and $w$ is a smooth, symmetric function
depending on the state. For a mathematically precise definition of
these quantities and of the statement that ``there are no spacelike singularities'', 
we refer to \cite{kaywald}. There exists an alternative, equivalent characterization of 
globally Hadamard states in terms of the wave front set (for 
a definition of this concept, see \cite{hor})  of their two-point function, 
due to Radzikowski~\cite{rad}, which plays a crucial r\^ole in the 
construction of normal orderd Wick products \cite{bfk} (given in the next subsection), and
of time ordered products \cite{bf} in curved spacetime. 
Namely, a state is Hadamard, if its two-point
function has the following wave front set:
\begin{eqnarray*}
\WF(\omega^{(2)}) = \{ (x_1, k_1, x_2, -k_2) \in 
(T^*M)^2 \backslash \{{\bf 0}\} \mid  (x_1, k_1) \sim (x_2, k_2), k_2 \triangleright 0\}.
\end{eqnarray*}
Here, the notation $(x_1, k_1) \sim (x_2, k_2)$ means that 
$x_1$ and $x_2$ can be joined by a null geodesic and that 
$k_1$ and $k_2$ are cotangent and coparallel to this geodesic. 
$k \triangleright 0$ means that $k_a$ is future pointing. 

\subsection{Review of Wick products of free fields in curved spacetime}
Brunetti, Fredenhagen and K\"ohler~\cite{bfk}  
gave a construction of Wick products of free fields in 
curved spacetime (with respect to some arbitrarily chosen
quasi-free reference Hadamard state $\omega$), 
which generalizes the well-known construction of Wick products in 
Minkowski space. These quantities are not already contained 
in the algebra $\cA$ of free fields and form the starting point 
for the perturbative construction of the interacting fields, 
which will be reviewed in the next subsection. 

To begin with, the authors of 
\cite{bfk} first define ``point-split Wick products'' by
\begin{equation*}
\lno \Phi(x_1) \dots \Phi(x_n) \rno_\omega \; =  
\frac{\delta^n}{i^n\delta f(x_1)
\dots \delta f(x_n)} {\rm exp}\left[\frac{1}{2}\omega^{(2)}(f, f)\myid 
+ i\Phi(f)
\right]
\Bigg|_{f=0}, 
\end{equation*}
where the field operators $\Phi$ on the right side of this equation 
mean the representers of the corresponding algebraic elements 
in the GNS-representation of $\omega$. (We note here that this 
normal ordering prescription is  equivalent to usual one defined in 
terms of creation and destruction operators. More precisely, 
suppose that the state $\omega$ in question had been obtained 
from some set $u_\lambda$ of positive frequency solutions 
to the Klein-Gordon equation. One could
then normal order the product $\Phi(x_1) \dots \Phi(x_n)$ 
by shifting all the creation operators $a^+_\lambda$ (corresponding 
to the modes $u_\lambda$) in 
that expression the left of all destruction operators. It is
not hard to see that the operator obtained in this way is
identical to the operator 
$\lno \Phi(x_1) \dots \Phi(x_n) \rno_\omega$ defined above.)  
\cite{bfk} demonstrated that the above Wick products are well-defined 
operator-valued distributions on $M^n= M \times \dots \times M$ ($n$ factors)
on a dense invariant domain $D_\omega$ in the GNS-Hilbert space $\cH_\omega$, and that they 
posses well-defined restrictions to all partial diagonals in $M^n$.
(A ``partial diagonal'' is a subset of $M^n$ of the form  
\begin{eqnarray*}
\Delta_{n_1, \dots, n_j}(M) = \{ (\underbrace{x_1, \dots, x_1}_{\text{$n_1$ times}}, \dots, 
\underbrace{x_j, \dots, x_j}_{\text{$n_j$ times}}) \mid x_i \in M, 
\; i = 1, \dots, j \}
\cong M^j.) 
\end{eqnarray*}
The restriction  of the point-split Wick product to 
such a partial diagonal is called 
a ``multi-local Wick product'' and is denoted by 
$\lno \Phi^{n_1}(x_1) \dots \Phi^{n_j}(x_j) \rno_\omega$. 
The restriction to the total diagonal, $\Delta_n(M)$, 
of the point-split Wick product is simply called a 
``Wick monomial'' and is denoted by 
$\lno \Phi^n(x) \rno_\omega$. 
We remark that while it is possible to define the point-split Wick products
for any quasifree state, this is not so for the above Wick monomials
(resp. multi-local Wick monomials). For the latter, it is crucial that 
one uses Hadamard states. 

We end this subsection by introducing some quantities and notions related 
to the above normal ordered Wick products. A linear combination of Wick monomials 
with coefficients in $\cD(M)$ is called a ``Wick polynomial''. Wick monomials and 
Wick polynomials involving derivatives
can be defined in a similar fashion from point-split 
Wick products of differentiated free fields. For any Wick-polynomial 
$A$ without derivatives, we define a Wick polynomial $\frac{\partial A}{
\partial \Phi}$ by 
$$
i \frac{\partial A}{\partial \Phi}(x) \Delta(x, y) = [A(x), \Phi(y)].
$$
For later purposes, we also find it convenient to define the degree, 
$\deg(A)$, of a Wick monomial $A$ as the number of free field factors 
plus the number of derivatives in $A$. 

\subsection{Review of perturbative interacting quantum field theory 
in curved spacetime}

The building blocks of perturbative interacting QFT   
(in the real-time formalism) are the time ordered products of Wick polynomials
(in curved spacetime, the latter are defined with respect to some arbitrary
reference Hadamard state $\omega$, which shall be kept fixed for the 
rest of this section). The time ordered products are
formally defined by
\begin{equation}
\label{TOP}
\text{``}\cT(A_1(x_1) \dots A_n(x_n)) = 
\sum_{{\rm perm}\,\pi} 
\vartheta(x_{\pi(1)}^0 - x_{\pi(2)}^0) \dots \vartheta(x_{\pi(n-1)}^0 - x_{\pi(n)}^0)
A_1(x_{\pi(1)}) \dots A_n(x_{\pi(n)}) 
\text{'',} 
\end{equation}
where $x^0$ is a global time-coordinate on $M$, 
$\vartheta$ is the distribution defined by 
\begin{eqnarray*}
\vartheta(t) = 
\begin{cases} 
\text{$1$ if $t \ge 0$,}\\
\text{$0$ otherwise,}
\end{cases}
\end{eqnarray*}
and $A_i, i = 1, \dots, n$ are Wick polynomials.  
Formula~\eqref{TOP} for the time ordered products cannot in 
general be taken at face value, since the operator products 
appearing on the right hand side of 
this equation are in general too (short-distance-) singular
to be multiplied by $\vartheta$--distributions, and therefore this equation is 
meaningless from a mathematical point of view.\footnote{
One aspect of this is that, 
if one naively `calculates' the above formal expression for the 
time ordered products (e.g. by applying what is usually called 
the ``naive Feynman rules'') then one is faced with ultraviolet
divergences which lead to meaningless infinite results.} 
There are however ways to construct mathematically well-defined, 
``renormalized'' operators $\cT(A_1(x_1) \dots A_n(x_n))$ 
which have the essential features of `time orderedness' formally 
shared by the ill-defined expression on the right hand side 
of Eq.~\eqref{TOP}. The process of constructing such 
time ordered products as well-defined operators on $\cH_\omega$
(with some common, dense and invariant domain) 
is usually referred to as ``renormalization''.  
This will be addressed in some more detail below; for 
the moment we assume that well-defined time ordered products
have been constructed, and proceed by explaining 
how one can locally define interacting quantum fields, algebras of 
observables etc. from these (renormalized) time ordered products. 

\medskip

The interaction of a Klein-Gordon field with itself is
described by an interaction Lagrangian $\cL(x) = \lambda A(x)$, 
where $\lambda$ is the coupling strength and $A$ is a 
Wick polynomial. In order to get a renormalizable theory 
$A$ must be a Wick polynomial of degree less or equal than four.\footnote{
In order for the interaction to be local, one must further 
require that $\cL(x)$ should be `locally defined in terms of the metric'
(for a precise mathematical formulation of this requirement see e.g.
\cite{wald}[item 2) on p. 89]). Such a requirement further reduces the possible form
of $\cL(x)$. Further restrictions come from dimensional considerations.} 
The corresponding $S$-matrix, $S[\cL]$, is defined, again formally, in 
terms of the renormalized 
time ordered products by 
the so-called ``Gell Mann--Low formula'' (viewed as a formal power series in 
the coupling strength $\lambda$), 
\begin{eqnarray}
\label{Smatrix}
\text{``}S[\cL] = \myid + \sum_{n=1}^\infty \frac{i^n}{n!}
\int \cT( \cL(x_1) \dots \cL(x_n) )
\,\mu_\g(x_1) \dots
\mu_\g(x_n) \text{''.}  
\end{eqnarray}
In Minkowski space, one obtains from the $S$-matrix---provided 
it exists (i.e., there is a suitable sense in 
which the above integrals converge at arbitrary order 
in perturbation theory)---the scattering amplitudes 
between asymptotic incoming and outgoing states. There are
however (otherwise perfectly reasonable) theories, 
for which the $S$-matrix does not properly 
exist, due to uncurable infrared divergences. Moreover,
in curved spacetimes which are not asymptotically flat in the remote past 
and future (for example, spacetimes with an initial big-bang singularity),
there is no well defined notion of a scattering process to begin with,  
regardless as to whether the above expression for the 
$S$-matrix makes good mathematical
sense or not. Although we think that for this reason the $S$-matrix does not 
occupy any fundamental status in a generic curved spacetime, it is still of
some value as a technical tool. For example, it can be used to 
construct the interacting theory locally on the algebraic level, as 
we shall briefly recall now (see ref.~\cite{bf} for details). 

For this purpose, one first introduces a local version of the 
$S$-matrix which is defined by the same formula as above, Eq.~\eqref{Smatrix}, 
but with the interaction Lagrangian replaced by the local quantity
$g(x)\cL(x)$ where $g$ is a 
smooth function of compact support. As a consequence the volume 
integrals in the expression for this local $S$-matrix now trivially converge.
For any Wick-polynomial $A(x)$ of free field operators 
one then introduces a corresponding interacting field operator (thought of 
as formal power series in $\lambda$) by 
\begin{eqnarray}
\label{intop}
A_{g\cL}(f) = 
S[g\cL]^{-1} \frac{\partial}{i\partial t} 
S[g\cL + tfA]
\Big|_{f=0}, \quad \text{$f \in \cD(M)$.} 
\end{eqnarray}
It is possible to expand the interacting field operators  
in terms of so-called ``totally retarded products'' (or $\cR$-products): 
\begin{eqnarray}
\label{Ret}
A_{g\cL}(y) = \sum_{n = 0}^\infty 
\frac{i^n}{n!} \int 
\cR(A(y) ; \cL(x_1) \dots \cL(x_n)) g(x_1) \dots g(x_n)\,
\mu_\g(x_1) \dots \mu_\g(x_n),  
\end{eqnarray}
where  
\begin{multline}
\label{TeRet}
\cR(A(y) ; A_1(x_1) \dots A_n(x_n)) = \\
\sum_{I \subset \{1, \dots, n\}} (-1)^{|I|} \cT\left(
\prod_{i \in I} A_i(x_i) \right)^+
\cT\left( A(y) \prod_{j \in I^c} A_j(x_j)  
\right).
\end{multline}
The $\cR$-products are symmetric in $x_1, \dots, x_n$ and have support 
in the set $\{(y, x_1, \dots, x_n) \mid x_j \in J^-(y), j =
1, \dots, n\}$. The commutator of two interacting fields 
is given by
\begin{eqnarray}
\label{comm}
&&[A_{1g\cL}(y_1), A_{2g\cL}(y_2)] = \sum_{n = 0}^\infty 
\frac{i^n}{n!} \int \mu_\g(x_1) \dots \mu_\g(x_n) g(x_1) \dots g(x_n)
\times \nonumber\\  
&& \Big( 
\cR(A_1(y_1) ; A_2(y_2)\cL(x_1) \dots \cL(x_n)) 
- \cR(A_2(y_2) ; A_1(y_1)\cL(x_1) \dots \cL(x_n)) \Big), 
\end{eqnarray}
see~\cite{df} for a proof of this formula.

We come to the definition of local algebras 
$\cA_\cL(\cO)$ of observables in the interacting theory~\cite{bf,df}. 
Let $\cO$ be a bounded open set in $M$ with compact closure. For any compactly 
supported, smooth function $g$ on $M$ such that $g \restriction \cO = 1$
(we denote the set of such functions by $\vartheta(\cO)$)
one first introduces the algebra
\begin{eqnarray*}
\fF_{g\cL}(\cO) = \langle A_{g\cL}(f) \mid f \in \cD(\cO), \quad 
\text{$A$ a Wick-polynomial} \rangle.  
\end{eqnarray*}
Now one can prove that if $g'$ is another function in $\vartheta(\cO)$,
then there exists a unitary $V$ such that 
\begin{eqnarray}
\label{intertw}
A_{g\cL}(f)V = VA_{g'\cL}(f) 
\end{eqnarray}
for all $f \in \cD(\cO)$ and any Wick polynomial $A$, thus 
showing that the algebraic structure of $\fF_{g\cL}(\cO)$ is actually 
independent of the choice of $g$. This observation leads one to the 
following definition for the algebras of observables $\cA_\cL(\cO)$: 
Let $\cV_{g, g'}(\cO)$ be the set of unitaries satisfying Eq.~\eqref{intertw}
for all $f \in \cD(\cO)$ and all Wick polynomials $A$. 
Then the local algebras $\fF_\cL(\cO)$ are by definition the algebras
generated by the set of all families 
\begin{equation}
a = (a_g)_{g\in\vartheta(\cO)}, \quad a_g \in \fF_{g\cL}(\cO),
\end{equation} 
with the property that $a_g V = V a_{g'}$ for all 
$V \in \cV_{g, g'}(\cO)$ and for any 
two functions $g, g'$ in $\vartheta(\cO)$. The product and $^*$--operation 
in this algebra are the obvious ones: 
\begin{eqnarray*}
(ab)_g := a_gb_g, \quad (a^*)_g := a_g^*.  
\end{eqnarray*}
By definition, 
the algebra $\fF_\cL(\cO)$ contains for example the elements
defined by $a = (A_{g\cL}(f))_{g \in \vartheta(\cO)}$, where $f \in 
\cD(\cO)$ and where $A$ is an arbitrary Wick polynomial. In the
following we denote such elements by $A_\cL(f)$. It is 
clear that if $\cO_1 \subset \cO_2$ are two spacetime regions with 
compact closure, then
every element in $\cA_\cL(\cO_1)$ can naturally be identified
with an element in $\cA_\cL(\cO_2)$, thus 
$\cA_\cL(\cO_1) \subset \cA_\cL(\cO_2)$. One may therefore 
define the algebra of all observables in the interacting
theory as the inductive limit, $\cA_\cL = \cup_\cO \cA_\cL(\cO)$. 
It follows from Eq.~\eqref{comm} and the 
support properties of the $\cR$-products that 
$$ [A_{1 \cL}(f_1), A_{2 \cL}(f_2)] = 0 $$
if the support of $f_1$ is spacelike related to the 
support of $f_2$. This means that the net $\cA_\cL(\cO)$ satisfies 
spacelike commutativity.

\medskip

In the previous paragraphs, we have indicated how quantities of 
interest in the interacting theory can be constructed from  
the time ordered products. In the remaining part of this subsection
we briefly address the issue of constructing the time ordered products
themselves. Our discussion follows \cite{bf}, who have generalized 
the methods of causal perturbation 
theory~\cite{eg,BS,stuck,S,stei2} in Minkowski space to 
general globally hyperbolic curved spacetimes. 
The main idea behind the causal approach is to construct 
the time ordered products by an inductive process  
based on their causal factorization properties. The induction proceeds 
by the order $n$ in perturbation theory
($=$ number of factors in the time ordered products). 
At first order, one sets $\cT(A(x)) := A(x)$ for any Wick
monomial $A$. The inductive assumptions 
on the time ordered products at a given order $n \ge 1$
(they are satisfied at first order) are the following:

\medskip
\noindent
{\bf Well-definedness:} 
The time ordered products are well-defined operator valued 
distributions on some dense, invariant domain $D_\omega \subset \cH_\omega$. 

\medskip
\noindent
{\bf Causal factorization:}
\begin{eqnarray*}
\cT(A_1(x_1) \dots A_n(x_n)) = \cT(A_1(x_1) \dots A_n(x_k)) 
\cT(A_1(x_{k+1}) \dots A_n(x_n)) 
\end{eqnarray*}
if 
\begin{eqnarray*}
x_j \notin J^-(x_i), \quad \text{for all $i = 1, \dots, k$ and
$j = k+1, \dots, n$,} 
\end{eqnarray*}
i.e., the time ordered products factorize if every point in 
one set of points is in the past of another set of points. 

\medskip
\noindent
{\bf Permutation symmetry:}
The time ordered product are symmetric under permutations of the
arguments.

\medskip
\noindent
{\bf Unitarity:}
\begin{equation}\label{N2}
\cT(A_1(x_1) \dots A_n(x_n))^+ = \sum_{{\mathcal P}}
(-1)^{n + |\cal{P}|} \prod_{I \in \cal{P}} \cT\left( \prod_{i \in I} A_i(x_i)^+\right),  
\tag{N2}
\end{equation}
where $\mathcal P$ runs over all partitions of $\{1, \dots, n\}$ into pairwise 
disjoint subsets. 

\medskip

{\it Remark:} This condition is equivalent to requiring that the $S$-matrix is unitary in 
the sense of formal power series of operators.

\medskip
\noindent
{\bf Wick expansion:}
Let $A_i, i = 1, \dots, n$ be Wick polynomials without derivatives. Then
\begin{eqnarray}
\label{causex}
\cT(A_1(x_1) \dots A_n(x_n)) = \sum_\alpha t^\alpha_n(x_1, \dots, x_n)/\alpha!\;
\lno \Phi^{\alpha_1}(x_1) \dots \Phi^{\alpha_n}(x_n) \rno_\omega,     
\end{eqnarray}
where
$$t^\alpha_n(x_1, \dots, x_n) = 
\left\langle \Omega_\omega \Bigg|
\cT\left(\frac{\partial^{\alpha_1} A_1}{\partial \Phi^{\alpha_1}}(x_1) \dots
\frac{\partial^{\alpha_n} A_n}{\partial \Phi^{\alpha_n}}(x_n) \right) \Omega_\omega \right\rangle.$$
(That the products between the various distributions in Eq. \eqref{causex}
indeed exist can be derived from the upper bound on the wave front 
set of the $t^\alpha_n$, given below, and a similar bound on 
the wave front set  of the Wick products. For details see
\cite[Microlocal Thm. 0]{bf}.) In our subsequent discussions, we will often use 
the following consequence of Eq. \eqref{causex}:
\begin{equation}
\label{N3}
[\cT(A_1(x_1) \dots A_n(x_n)),\Phi(y)] 
= \sum_{j=1}^n \cT\left(A_1(x_1) \dots
[A_j(x_j), \Phi(y)] \dots 
A_n(x_n)\right).  \tag{N3}
\end{equation}
Here, the time ordered products involving commutators on the r.h.s. are 
understood to mean the expression obtained by replacing $[A_j(x_j), \Phi(y)]$ with 
$i\Delta(x_j, y) \frac{\partial A_j}{\partial \Phi}(x_j)$ and by 
pulling the commutator functions out of the time ordered product. 

\medskip

{\it Remark:} An expansion formula similar to Eq. \eqref{causex}
holds also for time ordered products of Wick polynomials containing
differentiated field operators, it can e.g. be found in \cite{pr}. 
As for the case of undifferentiated Wick polynomials, this implies 
Eq. \eqref{N3}, where it is now understood that the expressions
$[A_j(x_j), \Phi(y)]$ are to be written in terms of suitably differentiated
commutator functions and suitable sub-Wick polynomials of $A_j$, 
see e.g. \cite{df,pr}.

\medskip
\noindent
{\bf Microlocal spectrum condition:}
The wave front set of the scalar time ordered distributions $t_n$ defined by 
\begin{equation}
\label{tndef}
t_n(x_1, \dots, x_n) = \langle\Omega_\omega |
\cT(A_1(x_1) \dots A_n(x_n))\Omega_\omega \rangle
\end{equation}
is contained in a certain set $\Gamma^{\cT}_n
\subset (T^*M)^n \backslash \{ {\bf 0} \}$, which is defined as follows: 
By definition, a point $(x_1, k_1; \dots; x_n, k_n) \in (T^*M)^n \backslash \{{\bf 0}\}$ 
is in $\Gamma^{\cT}_n$ if and only if (a) 
there exist null-geodesics $\gamma_1, \dots, \gamma_m$ which 
connect any point $x_j$ in the set $\{x_1, \dots, x_n\}$ to 
some other point in that set, 
(b) there exists coparallel, cotangent covectorfields $p_1, 
\dots, p_m$ along these geodesics such that $p_l \triangleright 0$
if the starting point of $\gamma_l$ is not in the causal 
past of the end point of $\gamma_l$, (c) for the covector 
$k_j$ over the point $x_j$ it holds that 
$k_j = \sum_e p_e(x_j) - \sum_s p_s(x_j)$, where the index
$e$ runs through all null-geodesics ending at $x_j$ and $s$ runs
through all null-geodesics starting at $x_j$. 

\medskip

{\it Remark:} The microlocal spectrum condition may be regarded as an analogue 
of the ususal spectrum condition in Minkowski spacetime. 

\medskip
\noindent
The next requirement asks that the distributions $t_n$, defined in 
Eq. \eqref{tndef},  have a certain scaling 
behaviour---roughly speaking, a certain ``asymptotic degree of homogeneity''---
at the total diagonal $\Delta_n(M)$ in $M^n$, which is 
compatible with the above microlocal spectrum condition. 
Mathematically, this can be expressed using the notion of 
the ``microlocal scaling degree'' $\mu\sd(t)$ of a distribution $t$, 
introduced in \cite{bf}. For a mathematically precise definition 
of this concept and some of its properties and related results, we refer to \cite{bf}.

\medskip
\noindent
{\bf Microlocal scaling degree:}
The distributions $t_n$, defined in Eq. \eqref{tndef}, 
have microlocal scaling degree 
\begin{equation}
\label{mlsd}
\mu{\rm sd}(t_n) \le \sum_j \deg(A_j)
\end{equation}
with respect to some cone $\Gamma_n$ (related to $\Gamma^\cT_n$) at the total 
diagonal $\Delta_n(M)$. In other words, the scaling behaviour of the 
time ordered products has to follow what is usually called
a ``power counting rule''.

\vspace{0.5cm}
\noindent
Let us assume now that time ordered products with the above properties 
have been constructed up to order $n$. Then, 
by the causal factorization property, it can be seen that those of order 
$n+1$ are already given as (operator-valued) 
distributions on the space $M^{n+1} \backslash \Delta_{n+1}(M)$, 
i.e., away from the total diagonal. The remaining task is thus to extend these
distributions to the entire space
$M^{n+1}$. It turns out that this is indeed possible and that 
the extension can be performed in such a way that the so-defined
time ordered products at order $n+1$ satisfy the 
requirements of causal factorization, permutation symmetry, 
unitarity, microlocal scaling degree and Wick expansion. 
However, the extension of these products to the total diagonal is in general 
not unique. Instead, there exist in general different extensions with 
the desired properties differing from each other by what is usually 
called a ``finite renormalization ambiguity''\footnote{
Of course finite renormalization ambiguities appear, in different guise,
also in all other renormalization procedures.}. 
The imposition of the above requirements on the 
time ordered products implies that this ambiguity must have a 
specific form. Firstly, by the 
Wick expansion property, the ambiguity in extending 
the time ordered products to the total diagonal 
is reduced to an ambiguity in extending 
the $\mc$-number distributions $t_{n+1}$. The scaling degree 
requirement then further reduces that
ambiguity to the choice of some set of smooth tensor fields on $M$. 
In order to make the latter statement more precise,   
we find it convenient to first introduce the notion of a ``symbol''. 
Let us denote points in 
$T^*_y M \times \dots \times T^*_y M$ ($n$ copies) by 
$(y, \underline k) = (y, k_1, \dots, k_n)$. 
\begin{defn}
A ``symbol'' $b$ is a map of the form 
\begin{eqnarray*}
b(y, \underline k) = \sum_{m = 0}^\rho \sum_{\pi \in {\bf P}(m,n)}
b_\pi^{a_1 \dots a_m}(y) k_{\pi(1) a_1} \dots k_{\pi(m) a_m}, 
\end{eqnarray*}
where ${\bf P}(m,n)$ is the set of all maps $\pi$ 
of the set $\{1, \dots, m\}$ to the set $\{1, \dots, n\}$,
and where the $b_\pi^{a_1 \dots a_m}$ are smooth tensor fields of 
rank $m$ over $M$. 
The order $\deg(b)$ of a symbol is defined as the maximum rank
of all nonzero tensor fields appearing in the above expression. 
The principal part $\sigma_b(y, \underline k)$ of a symbol $b$ is 
by definition the symbol obtained by collecting all terms in the 
above expression which contain nonzero tensor fields of rank $\deg(b)$. 
\end{defn}
\noindent
Any symbol $b$ defines a distribution $B \in \cD'(M^{n+1})$ by the formula
\begin{eqnarray}
\label{ambig}
B(y, \underline x) = b(y, -i\underline \nabla)\delta(y, \underline x). 
\end{eqnarray}
Here, we have used the abbreviations
$\underline\nabla = (\nabla^{x_1}, \dots, \nabla^{x_n})$, 
$\underline x = (x_1, \dots, x_n)$ as well as 
$\delta(y, \underline x) = \prod_{j = 1}^n \delta(y, x_j)$ with 
$\delta$ the covariant delta function on $M$. 
Conversely, any distribution $B$ supported on the total diagonal in 
$M^{n+1}$ and with microlocal scaling degree $\mu{\rm sd}(B) < \infty$ at the 
diagonal arises in the above form for a symbol $b$ of degree $\le \mu{\rm sd}(B) - 4n$. 

\medskip
\noindent
Now \cite{bf} showed that the imposition of the Wick expansion and the scaling requirements 
reduce the ambiguity in defining  $t_{n+1}(y, x_1, 
\dots, x_n)$ to a distribution supported on the total 
diagonal with a microlocal scaling degree given by Eq. \eqref{mlsd}. Therefore, by 
what we just said, it must be of the form Eq.~\eqref{ambig}, where $b$ is some 
symbol of degree $\le \mu{\rm sd}(t_{n+1}) - 4n$, whose form is further constrained by the unitarity
and symmetry requirements. In a renormalizable theory, the maximum microlocal scaling 
degree of the distributions $t_{n+1}$ appearing in the expansion for the 
time ordered products, Eq. \eqref{causex}, does not increase with the order 
$n$ in perturbation theory.
In the case at hand, these are the ones for which the interaction 
Lagrangian $\cL$ has degree less or equal than four. 

\medskip
\noindent
In this work, we need to impose a further normalization condition on the 
time ordered products (A proof of this for Minkowski space
is given in~\cite{stei2}, it can be adapted to curved spacetimes):

\medskip
\noindent
{\bf Equations of motion:}
\begin{multline}
\label{N4}
(\square_\g - m^2)_y \cT(A_1(x_1) \dots A_n(x_n) \Phi(y))\tag{N4}\\
=i\sum_{j = 1}^n \delta(y, x_j) \cT \left(A_1(x_1) \dots 
\frac{\partial A_j}{\partial \Phi}(x_j) \dots A_n(x_n) \right), 
\end{multline}
where we assume here, as will always be the case in the following, 
that the $A_j$ contain no derivatives. This condition ensures that
the interacting field $\Phi_\cL$ satisfies the non-linear 
equations of motion, in the sense that 
\begin{eqnarray*}
(\square_\g - m^2) \Phi_{g\cL}(x) = 
-g(\partial \cL/\partial \Phi)_{g\cL}(x),
\quad\text{for all $g \in \vartheta(\cO)$ and $x\in \cO$.}  
\end{eqnarray*} 

\section{Local implementation of spacetime symmetries for self-interacting 
theories on curved spacetime by Noether charges}
\label{sec2}
Let us suppose now that $\g_{ab}$  admits a nowhere vanishing Killing 
vectorfield $\xi^a$, i.e. $\pounds_\xi \g_{ab} = 
\nabla_{(a} \xi_{b)} = 0$, and let us 
denote by  $\psi^t: M \to M$ the flow generated by that Killing field.
The flow $\psi^t$ naturally induces a 1-parameter group of automorphisms $\alpha_t$ on the 
algebra $\cA$ of free quantum fields on $M$ by
$\alpha_t(\Phi(f)) = \Phi(f \circ \psi^{-t})$. 
Let us make the further assumption that there exists a quasifree 
Hadamard state 
$\omega$ on $\cA$ for which $\omega(\alpha_t(a)) = \omega(a)$ 
for all $a \in \cA, t \in \mr$, or equivalently, one for which 
\begin{eqnarray}
\psi^{t*} \omega^{(2)} = \omega^{(2)}\quad\text{for all $t \in \mr$.} 
\end{eqnarray}
Then there is a strongly continuous 1-parameter 
group of unitaries $\{U(t)\}_{t \in \mr}$ 
(with generator $H$) on the GNS-Hilbertspace $\cH_\omega$ which 
generates this flow on the field operators (in the 
GNS--representation corresponding to $\omega$), i.e., $\Phi(f
\circ \psi^{-t}) = U(t) \Phi(f) U(t)^+$ for all $f \in \cD(M)$. 

\subsection{The action of the symmetry group on the algebra of 
interacting fields}

Let us assume that the interaction Lagrangian $\cL$ is invariant under the 
spacetime symmetry, i.e., $\cL(f \circ \psi^{-t}) = 
U(t)\cL(f)U(t)^+$ for all $f \in \cD(M)$. Then 
it is possible to define an action of the 1-parameter group 
of spacetime symmetries $\{\psi^t\}_{t \in \mr}$ by 
$^*$-automorphisms $\alpha_{\cL,t}$ on the algebra 
of interacting fields which satisfy 
\begin{eqnarray*}
\alpha_{\cL,t}(\fF_\cL(\cO)) = \fF_\cL(\psi^t\cO).  
\end{eqnarray*}
for all bounded open regions $\cO \subset M$, where $\psi^t \cO$ denotes
the transported region. In order to construct these 
automorphisms, we need to impose a further normalization condition.

\medskip
\noindent
{\bf Covariance:}
Let $A_j, \, j = 1, \dots, n$ be Wick monomials. Then we demand that 
\begin{equation}
\label{xiinvar}
\cT(A_1(\psi^t(x_1)) \dots A_n(\psi^t(x_n))) = 
U(t) \cT(A_1(x_1) \dots A_n(x_n)) U(t)^+.
\end{equation}

\medskip

{\it Remark:} We will not give a full proof here that the covariance requirement 
can indeed be satisfied, but rather only sketch the main arguments. 
Firstly, the multilocal Wick products $\lno \Phi^{\alpha_1}(x_1)
\dots \Phi^{\alpha_n}(x_n) \rno_\omega$ by definition satisfy 
the covariance requirement, since the free field transforms as 
$\Phi(\psi^t(x)) = U(t)\Phi(x)U(t)^+$ and since the 
two-point function $\omega^{(2)}$
is by assumption invariant under $\psi^t$. In view of 
the expansion property, covariance of the time ordered products is 
therefore equivalent to $\psi^{t*} t_n = t_n$ for all $\mc$-number 
distributions $t_n$ of the form Eq.~\eqref{tndef}. 
As described in the preceeding section, the distributions
$t_{n}$ at a given order $n$ are obtained by an extension procedure 
from those at order $< n$. The crucial ingredient in 
that extension procedure \cite{bf} is a projection operator $W_n: \cD(M^{n})
\to \cD(M^{n})$ which maps any given testfunction to a testfunction
that vanishes on the total diagonal 
$\Delta_{n}(M)$ together with a fixed number 
of its derivatives (that number is actually equal to the singular degree of 
$t_n$, given by $\mu{\rm sd}(t_n) - 4(n-1)$). In order to prove the existence of an extension 
$t_{n}$ which 
is invariant under $\psi^t$, it is therefore sufficient to prove 
that there is a $\psi^t$-invariant choice for $W_n$ for all 
$n$, i.e., a choice for $W_n$, which commutes with the action of 
$\psi^{t*}$ on testfunctions, $W_n \circ \psi^{t*} = \psi^{t*} \circ W_n$. 
The crucial ingredient in the construction of 
$W_n$ (which potentially threatens the validity of this equation) is 
some bump function $w \in \cD(M \times M)$ which is 
equal to one in a neighbourhood 
of $\Delta_2(M)$. Now, using the fact that $\xi^a$ is by definition
nowhere vanishing, one can construct a bump function $w$ which is 
invariant under $\psi^t$, i.e., one for which $\psi^{t*} w = w$. 
It is not difficult to see that one can obtain from this a
$W_n$ which is invariant under $\psi^t$, leading thus 
to invariant distributions $t_n$ at every order in perturbation theory.  

\medskip

Now,  given an element $a \in \fF_\cL(\cO)$, i.e., a family
$a = (a_g)_{g\in\vartheta(\cO)}$ with $a_g \in \fF_{g\cL}(\cO)$ and the 
property that $Va_{g'} = a_g V$ for all $V \in \cV_{g, g'}(\cO)$, 
we define a family $(\alpha_{\cL, t}(a)_g)_{g\in \vartheta(\psi^{t}\cO)}$ 
of elements $\alpha_{\cL, t}(a)_g \in \fF_{g\cL}(\psi^{t}\cO)$ by 
\begin{eqnarray}
\alpha_{\cL,t}(a)_g = U(t) a_{g\circ
\psi^{t}} U(t)^+.
\end{eqnarray}
We first want to show that this map defines an element in 
$\fF_{\cL}(\psi^t\cO)$. This means that we have to verify that 
\begin{eqnarray}
V \alpha_{\cL, t}(a)_{g'} = \alpha_{\cL, t}(a)_{g} V
\end{eqnarray}
for all $V \in \cV_{g,g'}(\psi^t\cO)$. A moment of reflection 
shows that this is the case if $V(t):=U(t)^+V U(t)$ can be shown to be in
the intertwiner space $\cV_{g\circ \psi^t, 
g'\circ \psi^t} (\cO)$. In order to see the latter, let us take an 
arbitrary Wick polynomial $A$ and an $f \in \cD(\cO)$. Then, from the 
covariance requirement and the invariance of $\cL$, we get 
\begin{eqnarray*}
V(t)A_{g\circ\psi^t \cL}(f) &=& V(t)U(t)^+A_{g\cL}(f\circ\psi^{-t})U(t)\\
&=&U(t)^+VA_{g\cL}(f\circ\psi^{-t})U(t)\\
&=&U(t)^+A_{g'\cL}(f\circ\psi^{-t}) VU(t)\\
&=&A_{g'\circ \psi^t\cL}(f) V(t), 
\end{eqnarray*}
where in the second line we have used the expression \eqref{Ret} for 
the interacting field operators in terms of totally retarded products. 
Therefore $\alpha_{\cL, t}(a)$ is a well-defined element in $\fF(\psi^t\cO)$. 
It is clear from its definition 
that $\alpha_{\cL, t}$ respects the product and $^*$-operation 
in $\fF_\cL(\cO)$ and that 
$\alpha_{\cL, t+t'} = \alpha_{\cL, t} \circ \alpha_{\cL, t'}$. 
We have thus defined a 1-parameter group of  automorphism as claimed. 

Clearly, if there is no self-interaction, $\cL = 0$, 
the automorphism $\alpha_{\cL, t}$
coincides with the above defined automorphism $\alpha_t$ on the  
algebras of free fields. By definition, we also have that
\begin{eqnarray*}
\alpha_{\cL, t}(A_\cL(f)) = A_\cL(f \circ \psi^{-t})
\end{eqnarray*}
for any interacting operator $A_\cL(f)$ defined 
from a Wick polynomial $A$. The generator of the 1-parameter
group $\{\alpha_{\cL, t}\}_{t \in \mr}$ is given by the 
derivation 
\begin{equation*}
\delta_\cL(a) = i\frac{d}{dt} \alpha_{\cL, t}(a). 
\end{equation*}
The action of that derivation on 
elements of the form $A_\cL(f)$ it is 
given by $\delta_{\cL}(A_\cL(f)) = -iA_\cL(\xi^a \nabla_af)$.

\subsection{Implementation of $\delta_\cL$ by a local operator} 
The aim of this subsection is to show that for any
bounded region $\cO \subset M$, and any bounded, open, globally
hyperbolic region $\widehat\cO$ strongly containing $\cO$, there exists 
a charge operator $Q_\cL$ contained in $\cA_\cL(\widehat \cO)$, 
which implements the infinitesimal action $\delta_\cL$ of the 
spacetime symmetries on $\fF_\cL(\cO)$, 
\begin{eqnarray*}
\delta_\cL(a) = [Q_\cL, a], \quad\text{for all $a \in \cA_\cL(\cO)$,}
\end{eqnarray*}
and which arises, in some sense,  as a surface-integral
of an interacting Noether current-density corresponding to the 
symmetry $\xi^a$. As a preparation, we first consider the free-field case
(i.e., $\cL = 0$). 

\subsubsection{Free fields}

We consider the covariantly conserved\footnote{
It follows from the equations of motion that 
$\nabla^a \Theta_{ab} = 0$, therefore
$\nabla_a j^a = \nabla_a \Theta^{ab} \xi_b + 
\Theta^{ab}\nabla_{(a} \xi_{b)} = 0$.} 
free current-density $j^a = \Theta^{ab}\xi_b$, where   
\begin{eqnarray*}
\Theta_{ab} = \lno \nabla_a \Phi \nabla_b \Phi - \frac{1}{2}\g_{ab}
(\nabla^c \Phi \nabla_c \Phi + m^2 \Phi^2) \rno_\omega.  
\end{eqnarray*} 
We now construct from $j^a$ a free charge operator $Q$
in $\cA(\widehat \cO)$,  which implements the infinitesimal action of the 
symmetry on $\cA(\cO)$. In order to do so, we choose the following 

\medskip
\noindent
{\bf ``Technical data'':}
\begin{itemize}
\item
A  Cauchy surface $\widehat \Sigma$ in $\widehat \cO$
and Gaussian coordinates around $\widehat\Sigma$, 
described by the map $[-\epsilon, \epsilon]
\times \widehat \Sigma \owns (t, x) \mapsto \Exp_x tn^a \in M$, 
where $n^a$ is the surface normal of $\widehat \Sigma$.  
(Note that this implies that the vectorfield $n^a = (\partial/\partial t)^a$ 
satisfies $n^a n_a = -1$, $n^a\nabla_a n^b = 0$ and 
$\nabla_{[a} n_{b]} = 0$ near $\widehat \Sigma$.)  
\item
An open subset $\Sigma \subset \widehat \Sigma$ such that 
$\cO \subset D(\Sigma) \subset \widehat \cO$ and 
a function $f \in \cD(\widehat \cO)$ which can be written as
$f(x) = \rho(t(x))$ on a neighbourhood of $J(\cO) \cap \widehat \cO$, where 
$\rho$ is a smooth, real valued function with compact support on 
$[-\epsilon, \epsilon]$ such that $\int \rho(x^0)\, \rd x^0 = 1$, 
and $\epsilon$ is chosen so small that $\Exp_x tn^a$ is contained 
in $\widehat \cO$ for all $x \in \Sigma$ and $|t| \le \epsilon$.  
(Note that this implies that ${h^{a}}_b \nabla_a f = 0$ on $J(\cO) \cap \widehat \cO$, 
where $h^{ab} = \g^{ab} + n^a n^b$.)
\end{itemize}
We have the following Lemma:
\begin{lemma}
\label{freeQlem}
The operator 
\begin{equation}
\label{freeQ}
Q = j^a(n_a f) = \Theta^{ab}(n_a \xi_b f)
\end{equation} 
generates the Killing symmetry on all observables localized in $\cO$, $\delta(a)
= [Q, a]$ for all $a \in \cA(\cO)$ and for any $f, n^a$ as in ``technical data''. 
In other words, 
\begin{eqnarray*}
[Q, \Phi(x)] = i\xi^a\nabla_a \Phi(x) \quad\text{for all $x \in \cO$.}
\end{eqnarray*}
\end{lemma}
{\it Remark:} Note that $Q$ is morally given by an surface integral over a Cauchy 
surface of the time-component of $j^a$. This is clear because
the function $f(x)$ is of the form $\rho(t(x))$ on 
$J(\cO) \cap \widehat \cO$ for some bump function $\rho$, which 
one can imagine to be given by an ``infinitely sharp spike''.  
\begin{proof}
One computes
\begin{eqnarray*}
[Q, \Phi(x)] &=& \int_M [\Theta_{ab}(y), \Phi(x)] fn^a\xi^b(y)\,
\mu_\g(y)\\
&=& i \int_{J(\cO)} \Big( fn^{(a}\xi^{b)}(y)
\nabla_{a} \Phi(y) \nabla^y_{b} \Delta(x, y)
+ \nabla^{a}(fn^a\xi_b)(y) 
\Phi(y) \nabla^y_{a} \Delta(x, y) \Big) \,\mu_\g(y) 
\end{eqnarray*}
where in the second line we have performed a partial integration and
used the equations of motion. On the other hand
\begin{eqnarray*}
i\xi^a\nabla_a \Phi(x) &=& i\int_\Sigma \Phi(y)
\!\stackrel{\leftrightarrow}{\;\nabla^y_a} \xi^b \nabla^x_b 
\Delta(x, y) \,\rd S^a(y)\\
&=& i\int_{J(\cO)} \Big( \Phi(y) 
\!\stackrel{\leftrightarrow}{\;\nabla^y_a} \xi^b \nabla^x_b 
\Delta(x, y) \Big) n^af(y)\,\mu_\g(y)\\
&=& -i\int_{J(\cO)} \Big( \Phi(y)  
\!\stackrel{\leftrightarrow}{\;\nabla_a^y} (\xi^b \nabla^y_b 
\Delta(x, y)) \Big) n^af(y)\,\mu_\g(y) \\
\end{eqnarray*}
where in the second line we have used that $\xi^a\nabla^x_a
\Delta(x,y) = -\xi^a\nabla^y_a \Delta(x,y)$, which holds
because $\xi^a$ is a Killing field. This gives
\begin{multline*}
i\xi^a\nabla_a\Phi(x) = i \int_{J(\cO)} \Big(
fn^{(a}\xi^{b)}(y)\nabla_a \Phi(y) \nabla_b^y \Delta(x,y) \\ 
- (fn^b\nabla_b\xi^a(y) - 
   \nabla_b(fn^a\xi^b)(y))\Phi(y)\nabla^y_a \Delta(x,y)
\Big)\,\mu_\g(y), 
\end{multline*}
So, in order to prove the lemma, 
we must show that 
\begin{eqnarray*}
\nabla_b(fn^a\xi^b) - fn^b\nabla_b\xi^a = 
\nabla^a(fn^b\xi_b) 
\end{eqnarray*}
on $J(\cO) \cap \widehat \cO$.
The left side of this equation is equal to 
\begin{eqnarray*}
\text{l.s.} &=& 
fn^a \nabla_b \xi^b + (\xi^b \nabla_b f)n^a
+ \xi^b\nabla_b n^a f - n^b \nabla_b \xi^a f \\ 
&=& (\xi^b \nabla_b f)n^a + [\xi, n]^a f\\
&=& n^a
n^c \nabla_c f n^b \xi_b + 
n^a \xi^c {h_c}^b \nabla_b f
+ [\xi, n]^a f \\
&=& n^a n^c \nabla_c f n^b \xi_b 
+ [\xi, n]^a f, 
\end{eqnarray*}
where we have used that $\nabla_a \xi^a = 0$ by Killing's equation. 
The hand side is given by 
\begin{eqnarray*}
\text{r.s.} &=& \nabla^a f n^b \xi_b
+ f(\xi_b \nabla^a n^b + n_b \nabla^a \xi^b) \\
&=& \nabla^a f n^b \xi_b + [\xi, n]^a f\\
&=&n^a n^c \nabla_c f n^b \xi_b + [\xi, n]^a f, 
\end{eqnarray*}
where we have used that $\nabla^{(a} \xi^{b)} = 0$,    
$\nabla^{[a} n^{b]} = 0$ and that ${h^{a}}_b \nabla^b f = 0$ 
on $J(\cO) \cap \widehat\cO$. Hence both sides are equal, thus 
proving the lemma.
\end{proof}

\subsubsection{Interacting fields}

Let $f$ be a function of compact support in $\widehat\cO$
and  $n^a$ a timelike vectorfield, with the properties described 
in the previous subsection under ``technical data''. For any 
function $g$ in $\vartheta(\widehat\cO)$ we define an interacting current 
density by  
\begin{equation*}
J^a_{g\cL} = j^a_{g\cL} + g\xi^a\cL_{g\cL} = (\Theta_{g\cL}^{ab} + g\g^{ab}\cL_{g\cL})\xi_b,  
\end{equation*}
where the operator $j^a_{g\cL}$ is given 
by 
\begin{equation*}
j^a_{g\cL}(h_a) = S[g\cL]^{-1}\frac{\partial}{i\partial t}
S[g\cL + th_a j^a] = \Theta^{ab}_{g\cL}(h_a\xi_b),  
\end{equation*}
with $j^a$ the free current density.  
By analogy with the free charge, we
next define an interacting charge by 
$$Q_{g\cL} = J^a_{g\cL}(n_a f).$$ Our main 
result is that the interacting current density $J^a_\cL$ is 
conserved and the the corresponding charge operator $Q_\cL \in \fF_\cL(\widehat\cO)$ 
generates the symmetry induced by $\xi^a$ on the algebra $\fF_\cL(\cO)$. 
We formulate this as a theorem:
\begin{thm}
\label{mainthm}
The interacting current density $J^a_\cL$ is conserved, 
\begin{equation}
\label{stresscons}
\nabla_a J^a_{g\cL}(x) = 0 \quad 
\text{for all $x \in \cO$ and $g \in \vartheta(\widehat\cO)$,}
\end{equation}
in the sense of formal power series in the coupling constant
$\lambda$. The interacting charge $Q_{\cL}$ implements the 
infinitesimal action of the spacetime symmetry $\xi^a$ on 
interacting fields in the sense that 
$\delta_\cL(a) = [Q_\cL, a]$ for 
all $a \in \cA_\cL(\cO)$. In other words 
\begin{eqnarray}
\label{qgen}
[Q_{g\cL}, A_{g\cL}(x)] =  
i\xi^a\nabla_a A_{g\cL}(x) \quad\text{for all $x \in \cO$ and $g \in \vartheta(\widehat \cO)$,}  
\end{eqnarray}
and for any Wick polynomial $A$. 
(This holds for any choice of function $f$ and timelike vectorfield $n^a$ with the
properties described in the previous section under ``technical data''.) 
\end{thm}
\begin{proof}
The proof of the above result is divided into two parts. 
In the first part, we show that the time ordered products 
can be normalized in such a way that the interacting current density
$J^a_{g\cL}$ is covariantly conserved. A proof of this requires the 
demonstration of a corresponding set of Ward identities, see Eq. 
\eqref{wardid} below. A proof
of a similar set of Ward identities for the stress energy 
tensor in Minkowski space was previously given 
independently by \cite{pr} and \cite{ho}. Our proof of 
the Ward identities follows \cite{pr} and 
\cite{ho} closely, up to the point at which one has to remove the 
anomaly. Here the methods of the present paper differ from those 
of \cite{pr} and \cite{ho}, which are based on momentum space 
techniques and therefore not suitable in the present context. 
In the second part we then demonstrate (by a  
chain of arguments somewhat similar to the one given in 
\cite[pp 50--51]{pr}) that conservation
of the interacting current implies the second 
statement of the theorem, Eq. \eqref{qgen}. 

\medskip
\noindent
In order to show conservation of the interacting current, we
first expand
$J_{g\cL}^a$ in terms of totally retarded products (cf.~Eq.~\eqref{Ret}), 
and express these by products of time ordered products (cf.~Eq.~\eqref{TeRet}).  
It is then not difficult to see that Eq.~\eqref{stresscons} is equivalent to 
the set of Ward identities 
\begin{equation}
\label{wardid}
\nabla_a^y \cT(j^a(y)A_1(x_1) \dots A_n(x_n)) 
= i\sum_{j=1}^n \delta(y,x_j) \xi^a\nabla^{x_j}_a \cT( A_1(x_1) \dots \dots A_n(x_n) ), 
\end{equation} 
for all $n = 1, 2, \dots$ and all possible sub-Wick monomials 
$A_j, j = 1, \dots, n$ of $\cL$ (which therefore do not contain 
derivatives). A proof of the Ward identities, Eq. \eqref{wardid}, is 
given in the Appendix. 

\medskip
\noindent
We now show that the Ward identities imply Eq. \eqref{qgen}. 
Assume first that $J^-(x) \cap \supp(f) = \emptyset$. Then it is easy to see that 
there is a function $F \in \cD(\widehat \cO)$ for which $\nabla^a F = n^a f$ on 
some neighbourhood of $J^+(x) \cap \widehat \cO$ and for which $F = 1$ 
in some neighbourhood of $x$. We then have 
\begin{eqnarray*}
&&[j^a_{g\cL}(n_a f), A_{g\cL}(x)] = 
\sum_{n = 0}^\infty \frac{i^n}{n!} \int \mu_\g(\underline x) 
g(x_1) \dots g(x_n) \times \\
&&\int \mu_\g(y) n^a f(y) \Big( \cR(j^a(y); A(x) \underline \cL(\underline x)) - 
\cR(A(x); j^a(y) \underline \cL(\underline x)) \Big),  
\end{eqnarray*}
where we have used the abbreviation $\underline \cL(\underline x)
= \cL(x_1) \dots \cL(x_n)$. By the support properties of $f$ and the 
support properties of the totally retarded products, this is equal to 
\begin{eqnarray*}
&=& \sum_{n = 0}^\infty \frac{i^n}{n!} \int \mu_\g(\underline x) \mu_\g(y) 
g(x_1) \dots g(x_n) \nabla_a F(y) \cR(j^a(y); A(x) \underline \cL(\underline x)).  
\end{eqnarray*}
We next use the Ward identities for the $\cR$-products (which are 
easily obtained from the Ward-identities~Eq.~\eqref{wardid} for the 
$\cT$-products and formula Eq.~\eqref{TeRet}. This gives
\begin{eqnarray*}
&=& i\sum_{n = 0}^\infty \frac{i^n}{n!} \int \mu_\g(\underline x)\mu_\g(y)
g(x_1) \dots g(x_n)F(y)\delta(y, x)\xi^a\nabla^x_a \cR(A(x);\cL(\underline x))\\
&+& i\sum_{n = 1}^\infty \frac{i^n}{n!} \int \mu_\g(\underline x)\mu_\g(y)
g(x_1) \dots g(x_n)F(y)\sum_{j=1}^n \delta(y,x_j)\xi^a\nabla^{x_j}_a 
\cR(\cL(x_j);A(x)\underline \cL \backslash j(\underline x)),  
\end{eqnarray*}
where $\cL \backslash j(\underline x) = \cL(x_1) \dots \mslash \cL(x_j) 
\dots \cL(x_n)$. We observe that the first sum is just 
$i\xi^a\nabla_a A_{g\cL}(x)$. 
Performing the $x_j$-integrations in the 
second term, using that $\nabla^a F = n^a f$ on a neighbourhood of $J^+(x) \cap
\widehat \cO$, and observing that the $y$-support of $\cR(\cL(y); A(x) \dots)$
is contained in $J^+(x) \cap \widehat \cO$, we find that the 
above expression is equal to 
\begin{eqnarray*}
&=& i\xi^a\nabla_a A_{g\cL}(x) \\ 
&-&\sum_{n = 1}^\infty \frac{i^{n-1}}{n!} 
\sum_{j=1}^n
\int \mu_\g(x_1) \dots \mslash \mu_\g(x_j) \dots \mu_\g(x_n) \times \\
&& g(x_1) \dots \mslash g(x_j) \dots g(x_n)
n_a f(y) \cR(g\xi^a\cL(y);A(x)\underline \cL\backslash j(\underline x)).   
\end{eqnarray*}  
But the terms under the sum over $j$ are all equal, by the symmetry of 
the $\cR$-products, therefore this expression is  
(we shift the first summation index)
\begin{eqnarray*}
&=& i\xi^a\nabla_a A_{g\cL}(x) \\ 
&-&\sum_{n = 0}^\infty \frac{i^{n}}{n!} 
\int \mu_\g(\underline x)g(x_1) \dots g(x_n)
n_a f(y) \cR(g\xi^a\cL(y);A(x) \underline \cL(\underline x)).   
\end{eqnarray*}  
Now, again using that $\supp(f) \cap J^-(x) = \emptyset$ and the 
support properties of the totally retarded products, we see that 
we may add the term $\cR(A(x); g\xi^a\cL(y)\underline \cL(\underline x))$
under the integral, because this does not make any contribution. 
This then makes it obvious that the above expression is 
just 
$$ = i\xi^a\nabla_a A_{g\cL}(x) - [g\xi^a\cL_{g\cL}(n_a f), A_{g\cL}(x)],$$ 
which proves Eq.~\eqref{qgen} if
$\supp(f) \cap J^-(x) = \emptyset$. 
By a similar chain of arguments, one 
can prove that Eq.~\eqref{qgen} also holds if instead 
$\supp(f) \cap J^+(x) = \emptyset$. 
We will now show that the general case follows from these two facts. 
To this end, we will show that, for an arbitrary $f$ as in 
``technical data'', one can always construct a function 
$\widehat f$ with the same properties as $f$ and with the additional 
properties that: $\supp(\widehat f) \subset \supp(f)$, 
there holds either that $J^+(x) \cap \supp(\widehat f) = \emptyset$
or $J^-(x) \cap \supp(\widehat f) = \emptyset$ and 
\begin{eqnarray}
\label{div}
(f - \widehat f) n^a = \nabla^a F
\end{eqnarray}
in a neighbourhood of $J(x) \cap \widehat \cO$, 
where $F \in \cD(\widehat \cO)$. We may thus write
\begin{eqnarray*}
[J^a_{g\cL}(n_af), A_{g\cL}(x)] &=& [J^a_{g\cL}(n_a \widehat f), A_{g\cL}(x)] 
+ [J^a_{g\cL}(n_a(f - \widehat f)), A_{g\cL}(x)] \\
&=& i\xi^a\nabla_a A_{g\cL}(x) + [J^a_{g\cL}(\nabla_a F), A_{g\cL}(x)] \\
\end{eqnarray*}   
where in the last line we have used 
the fact that we already know Eq.~\eqref{qgen}
for functions like $\widehat f$. This then 
proves the theorem, because $J^a_{g\cL}(\nabla_a F) = 0$, by 
current conservation. 

It thus remains to construct 
an $\widehat f$ as in ``technical data'' in the previous 
subsection, such that in addition Eq.~\eqref{div} holds. Now recall 
that $f$ is of the form $f(x) = \rho(t(x))$ in a 
neighbourhood of $J(\cO) \cap \widehat\cO$ where 
$\rho$ is a compactly supported smooth function on 
$[-\epsilon,\epsilon]$ such that $\int \rho(y^0) \,\rd y^0 = 1$. 
Let us choose a $\widehat \rho$ with $\supp(\widehat \rho) \subset \supp(\rho)$ 
such that either $t(x) > \sup \, \supp(\widehat \rho)$ or
$t(x) < \inf \, \supp(\widehat \rho)$, and a function $\widehat f 
\in \cD(\widehat \cO)$ satisfying  
$\widehat f(x) = \widehat \rho(t(x))$ on $J(\cO) \cap \widehat \cO$. 
This function then clearly satisfies either 
$\supp(\widehat f) \cap J^+(x) = 0$ or 
$\supp(\widehat f) \cap J^+(x) = 0$. Let us define
\begin{eqnarray}
F(y) = \int_{-\infty}^{t(y)} \rho(y^0) - \widehat \rho(y^0) \, \rd y^0
\end{eqnarray}  
for $y \in J(\cO) \cap \widehat \cO$. It is clear that $F$ can be continued
smoothly to a function in $\cD (\widehat \cO)$. By definition
of $F$, Eq.~\eqref{div} holds in a neighbourhood of $J(x) \cap \widehat \cO$.
We have thus constructed a $\widehat f$ with the desired properties, thus 
finishing the proof.   
\end{proof}
 
\vspace{1cm}
\noindent
{\bf Acknowledgements:} I would like to thank K.~Fredenhagen for 
suggesting to me the idea for the proof of Eq.~\eqref{xiinvar}, as well as for pointing out several 
errors in earlier versions of this manuscript. This work was supported  
by NSF-grant PHY00-90138 to the University of Chicago.

\section{Appendix: Proof of the Ward identities, Eq. \eqref{wardid}}

\begin{proof}
We want to prove the Ward identities by an induction on $n$
and the total degree of the Wick monomials defined by
$\omega := \sum_{j=1}^n \deg(A_j)$. 
Let us write $D(y, x_1, \dots, 
x_n)$ for the left hand side minus the right hand side of 
Eq.~\eqref{wardid}, i.e., the anomaly and let us 
inductively assume that the 
Ward identities can be satisfied for some 
$n$ and some order $\omega$. The logic of the induction step is the
following. In Step~1), we show that the Ward identities for 
$\omega + 1$ and $n$, can be reduced to the scalar identity obtained 
by taking the vacuum expectation value of Eq.~\eqref{wardid}.
It is argued in Step~3) that these can be satisfied. By what we have
just said, it 
is then sufficient to show the Ward identities for $(n + 1)$, $\rho$, 
when one of the $A_j$ is equal to $\Phi$. 
Again, arguing as in Step~1), only the
scalar identity has to be proven. This is done in Step~2). 

\medskip

{\bf Step 1}:
We  want to show that $D(y, x_1, \dots, x_n)$
is a multiple of the identity operator. 
To show this, we first demonstrate
that it commutes with any free field operator. We have
\begin{eqnarray*}
&&[D(y, \underline x), \Phi(z)] =
\nabla_a^y\cT\left([j^a(y),\Phi(z)] 
A_1(x_1) \dots A_n(x_n)\right) +\\
&& 
i\sum_{j=1}^n \Delta(x_j,z) \, \nabla_a^y 
\cT\left( j^a(y)
\dots \frac{\partial A_j}{\partial \Phi} (x_j) \dots A_n(x_n)
\right) + \\
&&
i\sum_{j=1}^n \delta(y,x_j)\xi^a\nabla^{x_j}_a 
\sum_{k=1}^n\Delta(x_k,z) 
\cT\left( A_1(x_1) \dots \frac{\partial A_k}{\partial \Phi}(x_k) \dots A_n(x_n)\right).
\end{eqnarray*}
To proceed, we calculate
\begin{equation*}
[j^a(y),\Phi(z)]= 
i\Big( \nabla^{(a}\Phi(y) \nabla^{b)}\Delta(y, z) 
-\g^{ab}(\nabla^c \Phi(y)\nabla_c\Delta(y, z) 
+ m^2 \Phi(y)\Delta(y, z))\Big)\xi_b(y). 
\end{equation*}
We now demand that time ordered products containing a once differentiated 
free-field factor satisfy the following normalization condition\footnote{
This condition is essentially equivalent to a generalization of 
(N4) to time ordered products containing a once differentiated
free field factor. For a more detailed discussion, see \cite[p. 52]{pr}.}
$$
\cT( \nabla_a \Phi(y) A_1(x_1) \dots A_n(x_n))
= \nabla_a^y\cT( \Phi(y) A_1(x_1) \dots A_n(x_n)).
$$
The above expression for $[j^a(y), \Phi(z)]$ together 
with the normalization condition  (N4) then gives:
\begin{multline}
\label{form1}
\nabla^y_a \cT\left([j^a(y),\Phi(z)] 
A_1(x_1) \dots A_n(x_n)\right) = \\
i\sum_{j = 1}^n \delta(x_j,y) \xi^a\nabla_a \Delta(y,z) \,
\cT\left( A_1(x_1) \dots \frac{\partial A_j}{\partial \Phi}(x_j)
\dots A_n(x_n) \right).
\end{multline}
From this, we obtain
\begin{multline*}
[D(y, \underline x), \Phi(z)] 
= i\sum_{j=1}^{n} \Delta(x_j,z)\Bigg\{
\nabla^y_a
\cT\left( j^a(y)
\dots \frac{\partial A_j}{\partial \Phi} (x_j) \dots A_n(x_n)
\right)\\ 
-i\sum_{k=1}^n \delta(y,x_k)\xi^a\nabla^{x_k}_a
\cT\left( A_1(x_1) \dots \frac{\partial A_j}{\partial \Phi}(x_j) \dots A_n(x_n)\right)
\Bigg\}.
\end{multline*}
Now the expression in braces vanishes by the Ward identities at 
total degree less than $\omega$, showing thus that $D$ commutes with 
the free field. By the Wick expansion requirement, the 
operator $D$ must be a linear combination of (multi-local) 
Wick products with distributional coefficients. Now it can be 
seen that any operator of this form which commutes 
with a free field is in fact  a multiple of the identity, 
showing thus that $D$ is a $\mc$-number. 

We next want to show that the numerical distribution
$D$ is localized at $x_1 = \dots = x_n = y$. In order to see this, 
note that for any point $M^{n+1} \owns
(x_1, \dots, x_n, y) \neq (y, \dots, y)$, one can find a 
Cauchy surface $\Sigma$ in $M$ which separates some points, 
$\{x_j\}_{j \in J, J \neq \emptyset}$ say, from the other
points $\{x_j\}_{j \notin J}$ and $y$. Without loss of generality
we assume that the latter 
are in not in the causal past of the first set of points. 
Then, by causal factorization and the induction hypothesis, we have
\begin{eqnarray*}
D(y, x_1, \dots, x_n) = 
\underbrace{D(y, \{x_j\}_{j \notin J})}_{=0}\, 
\cT\left(\prod_{j \in J} A_j(x_j)\right) = 0.
\end{eqnarray*}
Therefore we conclude $D(y, \underline x)$ is a scalar distribution 
supported on the total diagonal in $M^{n+1}$. 

\medskip

{\bf Step 2:} We show that the scalar Ward identities hold for 
$n+1$ factors when one of the
Wick monomials is a free field. We find from the formula for the 
time ordered products with a free field factor, Eq.~\eqref{N4}, that 
\begin{eqnarray*}
&&\left\langle \Omega_\omega| 
\cT(j^a(y)\Phi(x)A_1(x_1) \dots A_{n}(x_{n})) 
\Omega_\omega \right\rangle \\ 
&=&i\sum_{j=1}^{n} \Delta_F(x,x_j) 
\left\langle\Omega_\omega \Bigg|
\cT\left(j^a(y)A_1(x_1) \dots \frac{\partial A_j}{\partial \Phi}
(x_j)\dots A_{n}(x_{n})\right) 
\Omega_\omega\right\rangle\\ 
&+&i\xi_b(y)\nabla^{(b} \Delta_F(y,x) \left\langle\Omega_\omega|
\cT\left(\nabla^{a)}\Phi(y)A_1(x_1) \dots A_{n}(x_{n})\right) 
\Omega_\omega\right\rangle\\
&-&i\xi^a(y)(\nabla_b\Delta_F(y,x)\left\langle\Omega_\omega,
\cT\left(\nabla^b \Phi(y)A_1(x_1) \dots A_{n}(x_{n})\right) 
\Omega_\omega\right\rangle\\
&+&m^2\Delta_F(y,x)\left\langle\Omega_\omega |
\cT\left( \Phi(y)A_1(x_1) \dots A_{n}(x_{n})\right) 
\Omega_\omega\right\rangle) 
\end{eqnarray*}
where $\Delta_F= \omega^{(2)} + i\Delta_A$ 
is the Feynman propagator. From this one obtains
\begin{eqnarray*}
&&
\nabla^y_a \left\langle\Omega_\omega|
\cT(j^a(y)\Phi(x)A_1(x_1) \dots A_{n}(x_{n})) 
\Omega_\omega\right\rangle \\
&=&
i\sum_{j=1}^{{n}} \Delta_F(x, x_j) 
\nabla^y_a \left\langle\Omega_\omega \Bigg|
\cT\left(j^a(y)A_1(x_1) \dots \frac{\partial A_j}{\partial \Phi}
(x_j)\dots A_{n}(x_{n})\right) 
\Omega_\omega\right\rangle\\
&+&
i\xi^a\nabla_a\Delta_F(y,x) (\square_\g - m^2)_y \left\langle\Omega_\omega|
\cT\left(\Phi(y)A_1(x_1) \dots A_{n}(x_{n})\right) 
\Omega_\omega\right\rangle \\
&+&
i\delta(y,x)\left\langle\Omega_\omega|
\cT\left(\xi^a\nabla_a\Phi(y)A_1(x_1) \dots A_{n}(x_{n})\right) 
\Omega_\omega\right\rangle. 
\end{eqnarray*}
Now using condition \eqref{N4} and the Ward identities for 
$n$ factors, one finds that this is equal to 
\begin{eqnarray*}
&=&-\sum_{j,k=1}^{{n}}
\Delta_F(x,x_j) \delta(y,x_k)\xi^a\nabla^{x_k}_a\left\langle\Omega_\omega\Bigg|
\cT\left(A_1(x_1) \dots \frac{\partial A_j}{\partial \Phi}
(x_j)\dots A_{n}(x_{n})\right) 
\Omega_\omega\right\rangle\\
&-&
\sum_{j=1}^{n} \xi^a \nabla_a \Delta_F(y,x) \delta(y,x_j)
\left\langle\Omega_\omega \Bigg|
\cT\left(
\Phi(y)A_1(x_1) \dots \frac{\partial A_j}{\partial \Phi}
(x_j) \dots A_{n}(x_{n})
\right) 
\Omega_\omega\right\rangle\\
&+&
i\delta(y,x)\xi^a\nabla^y_a\langle\Omega_\omega|
\cT(\Phi(y)A_1(x_1) \dots A_{n}(x_{n})) 
\Omega_\omega\rangle\\
&=&
i\sum_{j=1}^{n} \delta(y,x_j) \xi^a\nabla^{x_j}_a 
\left\langle\Omega_\omega|
\cT\left(
\Phi(x) A_1(x_1) \dots A_{n}(x_{n}) 
\right)
\Omega_\omega\right\rangle\\
&+& 
i\delta(y,x)\xi^a\nabla_a^x \left\langle\Omega_\omega | 
\cT\left(\Phi(x)A_1(x_1) \dots A_{n}(x_{n})\right) 
\Omega_\omega\right\rangle.  
\end{eqnarray*}
Hence we have shown that the scalar Ward identities hold for $n$ factors 
if one of the factors is a free field. 

\medskip

{\bf Step 3):} We now show that one can remove the anomaly
by a suitable redefinition of the time ordered products. In 
order to do this, we first show that 
it is possible to write the $\mc$-number distribution 
$D$ as the total divergence 
\begin{eqnarray}\label{totdiv}
D(y, x_1, \dots, x_n) = \nabla^y_a D^a(y, x_1, \dots, x_n), 
\end{eqnarray} 
of some vector-valued distribution $D^a$ which is supported on the 
total diagonal of $M^{n+1}$ having microlocal scaling 
degree $\mu\sd(D) - 1$ on the 
total diagonal and which is invariant under the flow $\psi^t$. 
It is clear that the redefined time ordered products 
$\cT(j^a A_1 \dots A_n) - D^a \cdot \myid$ then satisfy the Ward identities 
and all the other requirements, thus concluding the proof. 

Let us call a symbol $b$ ``invariant''
if there holds $b(\psi^t(y), \underline k)
= b(y, \psi^{t*} \underline k)$ for all $t\in\mr$. In addition to 
scalar valued symbols, we also want to consider
tensor-valued symbols $b^{a_1 \dots a_s}$. These 
are defined in the same way as the ordinary symbols above, but with
the difference that $b^{a_1 \dots a_s}(y, \underline k)$ is 
now a tensor in the tangent space at $y$. The rank and 
the principal part of such symbols are defined by analogy to 
the scalar case. An invariant, tensor-valued symbol is one for 
which $b^{a_1\dots a_s}(\psi^t(y), \underline k)
= \psi^t_* b^{a_1 \dots a_s}(y, \psi^{t*} \underline k)$ for all $t$.  
Any contraction or tensor product with $\g_{ab}, \g^{ab}, \xi_a, k_a$ or covariant 
derivative with respect to $y$ of an invariant, 
tensor-valued symbol gives again a symbol of that kind. 
Any distribution $B \in \cD'(M^{n+1})$ which is supported on the 
total diagonal with microlocal scaling degree $\mu{\rm sd}(B) < \infty$ 
(w.r.t. the total diagonal)  and 
which is addition invariant under the flow $\psi^t$ (such as for example $D$) 
arises in the form \eqref{ambig} from an invariant symbol $b$ of degree
$\mu{\rm sd}(B) - 4n$, and vice-versa. 
Analogous statements hold true for tensor-valued symbols.

Let $d$ be the invariant symbol corresponding to $D$. Let us assume  
that there exist vector-valued, invariant symbols $d^a$ with the
property that 
\begin{equation}
\label{propdiv}
\nabla^y_a d^a(y, \underline k) + 
d^a(y, \underline k) \sum_{j = 1}^n 
ik_{j a} = d(y, \underline k)
\end{equation}
for all $(y, \underline k) \in T^*_y M \times \dots \times T^*_y M$, 
and let $D^a$ be the corresponding vector-valued, invariant distribution
on $M^{n+1}$. Then $D^a$ satisfies Eq.~\eqref{totdiv}. It thus remains to 
construct a vector valued, invariant symbol $d^a$ with the property 
Eq.~\eqref{propdiv}. In order to do that, we start with a lemma.
\begin{lemma}
The distribution $D$ (the anomaly) has the property
$D(1 \otimes f) = 0$ for all $f \in \cD(M^n)$, in other words
\begin{equation*}
\int_M D(y, \underline x)\,\mu_\g(y) = 0.
\end{equation*}
\end{lemma}
\begin{proof}
Let us chose a bounded region $\cO$ containing the points 
$x_1, \dots, x_n$ and a function $F$ which is equal to one on a 
neighbourhood of
$\cO$ and which has the property that $\nabla_a F = n_a f_1 - n_a f_2$ 
where $f_1$ and $f_2$ and $n^a$ are as under ``Technical Data'' in 
the previous section (with $\widehat \cO$ taken to be $M$) and where 
in addition $\supp(f_1) \cap J^+(\cO) = \emptyset$ and $\supp(f_2) 
\cap J^-(\cO) = \emptyset$. Then by causal factorization, 
\begin{eqnarray*}
\int D(y, \underline x) F(y) \,\mu_\g(y)  
&=& j^a(f_1n_a) \cT(A_1(x_1) \dots A_n(x_n)) 
- \cT(A_1(x_1) \dots A_n(x_n))j^a(f_2 n_a) \\
&-&i\sum_{j=1}^n \xi^a\nabla^{x_j}_a \cT( A_1(x_1) \dots A_n(x_n))\\
&=&[j^a(f_1n_a), \cT(A_1(x_1) \dots A_n(x_n))] 
- \cT(A_1(x_1) \dots A_n(x_n)) j^a(\nabla_a F)\\
&-&i\sum_{j=1}^n \xi^a\nabla^{x_j}_a \cT( A_1(x_1) \dots A_n(x_n)).
\end{eqnarray*}
The second to last term vanishes by current conservation. 
Now by Lem.~\ref{freeQlem}, we have $[j^a(f_1n_a), \Phi(x)] 
= [H, \Phi(x)]$ for all $x \in \cO$. Together with the 
Wick expansion requirement on the time ordered 
products, Eq. \eqref{causex}, one can conclude from this that  
\begin{eqnarray*}
\left[j^a(f_1n_a), \cT(A_1(x_1) \dots A_n(x_n)) \right] =
\left[H, \cT(A_1(x_1) \dots A_n(x_n)) \right],
\end{eqnarray*}
where $H$ is the generator of the group $U(t)$. 
Hence it remains to show that
\begin{equation}
\label{donald}
\left[H, \cT(A_1(x_1) \dots A_n(x_n)) \right] 
-i\sum_{j=1}^n \xi^a\nabla^{x_j}_a\cT \left( A_1(x_1) \dots A_n(x_n) \right) = 0.
\end{equation}
But this equation is just the infinitesimal version of the 
covariance requirement, so the lemma is proven. 
\end{proof}
The lemma says that
\begin{equation}
\label{master}
\int_M [d(y, -i\underline \nabla)f](y, \dots, y)
\,\mu_\g(y)= 0 \quad\text{for all $f \in \cD(M^n)$}.
\end{equation}
We are now going to show that this equation implies the existence
of vector-valued, invariant symbols $d^a$ satisfying Eq.~\eqref{propdiv}. 
We do this by an induction 
in the degree $\rho$ of $d$. If $\rho = 0$, then it is easy 
to see that Eq.~\eqref{master} already implies $d = 0$, so in that
case Eq.~\eqref{propdiv} can trivially be satisfied by choosing $d^a = 0$.  
Let us now assume that Eq.~\eqref{master} had been shown to imply the existence
of vector-valued, invariant symbols $d^a$ satisfying Eq.~\eqref{propdiv}, whenever
the degree of $d$ is less or equal to $\rho - 1$. We need to 
show that we can construct symbols $d^a$ with 
the desired properties also if the degree of $d$ is $\rho$. 

To do this, we want to exploit Eq.~\ref{master} by testing it with 
compactly supported functions $f$ of the form $\chi \otimes \eta$, where 
$\chi = \chi(x_1)$ and $\eta = \eta(x_2, \dots, x_n)$ are 
testfunctions in $\cD(M)$ and $\cD(M^{n-1})$ respectively. 
Let us expand $d$ in terms of $k_1$, 
\begin{eqnarray*}
d = 
d^{(0)a_1 \dots a_\rho} k_{1 a_1} \dots k_{1 a_\rho} + 
d^{(1)a_1 \dots a_{\rho-1}} k_{1 a_1} \dots k_{1 a_{\rho - 1}}
+ \dots +
d^{(\rho)}
\end{eqnarray*}
where $d^{(m)a_1 \dots a_{\rho - m}} = 
d^{(m)a_1 \dots a_{\rho - m}} (y, k_2, \dots, k_n)$
are tensor-valued, invariant symbols of degree $m$, depending on the indicated arguments.
From Eq.~\eqref{master} with $f = \chi \otimes \eta$ we conclude that
\begin{eqnarray*}
0 &=& \sum_{m = 0}^\rho (-i)^{\rho - m} \int_M 
\nabla_{a_1} \dots \nabla_{a_{\rho - m}} \chi(y) 
[d^{(m)a_1 \dots a_{\rho - m}}(y, -i\nabla^{x_2}, \dots, -i\nabla^{x_n})
\eta](y, \dots, y) \,\mu_\g(y) \\
&=& \sum_{m = 0}^\rho i^{\rho - m} \int_M 
\chi(y) 
\nabla_{a_1}^y \dots \nabla_{a_{\rho - m}}^y
[d^{(m)a_1 \dots a_{\rho - m}}(y, -i\nabla^{x_2}, \dots, -i\nabla^{x_n})
\eta](y, \dots, y) \,\mu_\g(y) \\
&=:& \int_M \chi(y) [p(y, i\nabla^{x_2}, \dots, i\nabla^{x_n})\eta]
(y, \dots, y) \,\mu_\g(y), 
\end{eqnarray*}
where the symbol $p$ is defined by the last equation. This implies that
\begin{equation*}
[p(y, -i\nabla^{x_2}, \dots, -i\nabla^{x_n})\eta](y, \dots, y) = 0 
\quad\text{for all $y \in M$ and $\eta \in \cD(M^{n-1})$.}
\end{equation*}
It follows from this that the symbol $p$ vanishes (and hence also its
principal symbol), in other words
\begin{multline*}
0 = \sigma_p(y, k_2, \dots, k_n) = \\
\sum_{m = 0}^\rho \left[\sigma_{d^{(m)}}
(y, k_2, \dots, k_n)^{a_1 \dots a_{\rho - m}} \sum_{j_1, \dots, j_{\rho - m} \neq 1}
(-1)^{\rho - m}k_{j_1a_1} \dots k_{j_{\rho - m}a_{\rho - m}}
\right]. 
\end{multline*}
The point is now that the right side of this equation is actually  
$$=\sigma_{d}(y, k_1 = -k_2 - \dots -k_n, k_2, \dots, k_n),$$ which shows that 
\begin{eqnarray}
\sigma_{d}(y, \underline k) = b^a(y, \underline k)\sum_{j=1}^n k_{ja}
\end{eqnarray}
for some vector-valued, invariant symbol $b^a$ of degree less or equal to 
$\rho - 1$.  Let us now set
$$ r(y, \underline k) := d(y, \underline k) -
\nabla^y_a b^a (y, \underline k) - 
b^a (y, \underline k )\sum_{j=1}^n ik_{ja}.$$ 
$r$ is by definition an invariant, vector-valued symbol of degree 
$\le \rho - 1$. By construction, $r$ satisfies
satisfies Eq.~\eqref{master}. Thus we can apply the induction hypothesis
and conclude that there are invariant, vector-valued symbols $r^a$ of 
degree $\le \rho -2$ such that
\begin{equation*}
\nabla^y_a r^a(y, \underline k) + 
r^a(y, \underline k) \sum_{j = 1}^n 
ik_{ja} = r(y, \underline k).
\end{equation*}
From this one immediately concludes that 
the invariant, vector-valued symbol  $d^a = r^a + b^a$ 
satisfies Eq.~\eqref{propdiv}, thus concluding the proof.
\end{proof}

\end{document}